\documentclass[11pt,a4paper]{article}
\usepackage[margin=1in]{geometry}
\usepackage{authblk}
\usepackage[utf8]{inputenc}
\usepackage[T1]{fontenc}
\usepackage{url}
\usepackage{booktabs}
\usepackage{amsfonts}
\usepackage{amsmath}
\usepackage{nicefrac}
\usepackage{microtype}
\usepackage{xcolor}
\usepackage{graphicx}
\usepackage{enumitem}
\usepackage{multirow}
\usepackage{tcolorbox}
\usepackage{tikz}
\usetikzlibrary{arrows.meta, shapes.geometric, positioning, fit, backgrounds, calc, decorations.pathreplacing}
\usepackage{natbib}
\usepackage{hyperref} %

\tcbset{
  graybox/.style={colback=gray!5, colframe=gray!50, boxrule=0.5pt, arc=2pt,
    left=4pt, right=4pt, top=2pt, bottom=2pt, fontupper=\small}
}

\begin{document}

\title{Reflect-Guard: Enhancing LLM Safeguards against Adversarial Prompts via Logical Self-Reflection}

\author[yale]{Lixing Lin}
\author[ind]{Juli You}
\author[columbia]{Yue Li}
\author[ind1]{Luyun Lin}
\author[ind1]{Yiqing Wang}
\author[ind]{Zhen Zhang}
\author[ind]{Moxuan Zheng}
\affil[yale]{Yale University, New Haven, CT, USA}
\affil[ind]{Independent Researcher}
\affil[columbia]{Columbia University, New York, NY, USA}
\affil[ind1]{Citigroup, Dallas, USA}

\date{}

\maketitle

\begin{abstract}

Large language model (LLM) safety classifiers such as Llama Guard are effective at detecting overtly harmful prompts but remain vulnerable to adversarial jailbreak attacks that disguise malicious intent through role-play scenarios, fictional framing, and indirect requests. We present \textbf{Reflect-Guard}, a method that augments LLM-based safety classifiers with chain-of-thought self-reflection capabilities through parameter-efficient fine-tuning. Our approach distills analytical reasoning from GPT-4o-mini into structured reflection annotations, then trains Llama-Guard-3-8B via QLoRA to generate logical self-reflections before issuing safety verdicts. Using only 1{,}000 training examples and updating just 0.5\% of model parameters ($\sim$42M), Reflect-Guard achieves substantial improvements on two challenging benchmarks. On WildGuardTest, F1 score improves from 0.770 to 0.842 (+7.2 pp), with recall on adversarial prompts increasing from 0.513 to 0.921 (+40.8 pp). On JailbreakBench, the attack success rate drops from 10.3\% to 1.8\%, representing an 82.5\% relative reduction. These gains are especially pronounced on adversarial inputs, where the explicit reasoning step enables the model to see through obfuscation techniques that defeat standard pattern-matching approaches. Our results demonstrate that teaching safety classifiers to reason about adversarial intent, rather than simply classify surface patterns, is a promising direction for robust LLM safety.

\end{abstract}


\section{Introduction}

The deployment of large language models (LLMs) in production systems has created an urgent need for robust safety classifiers that can intercept harmful requests before they reach the target model~\citep{inan2023llamaguard, han2024wildguard}. Current safety classifiers, including Meta's Llama Guard family~\citep{inan2023llamaguard, metallamaguard3}, perform well on straightforward harmful prompts but exhibit significant vulnerabilities when confronted with adversarial jailbreak attacks~\citep{chao2023jailbreaking, zou2023universal, liu2024autodan}.

Adversarial jailbreaks exploit a fundamental weakness in pattern-based classification: by wrapping harmful requests in benign-sounding contexts---fictional narratives, role-play scenarios, hypothetical research questions, or educational framings---attackers can bypass safety filters that rely primarily on surface-level semantic matching~\citep{wei2024jailbroken}. For instance, Llama-Guard-3-8B achieves only 51.3\% recall on the adversarial subset of WildGuardTest, meaning it misses nearly half of disguised harmful prompts.

We hypothesize that this vulnerability stems from the classifier's lack of explicit reasoning about adversarial intent. Human safety reviewers naturally decompose ambiguous inputs: they identify framing techniques, assess underlying intent behind surface-level language, and reason about whether a request would cause harm regardless of its narrative wrapper. Current classifiers lack this analytical capability, making binary decisions without interpretable intermediate reasoning.

We propose \textbf{Reflect-Guard}, a method that endows safety classifiers with chain-of-thought (CoT) self-reflection capabilities through knowledge distillation and parameter-efficient fine-tuning. Our approach proceeds in three stages: (1) we use GPT-4o-mini to generate structured reflection annotations for a curated set of 1{,}000 training examples, analyzing adversarial techniques and harm indicators; (2) we fine-tune Llama-Guard-3-8B using QLoRA~\citep{dettmers2024qlora} to generate these reflections as an intermediate reasoning step before safety classification; and (3) at inference time, the model produces an explicit logical analysis in \texttt{<reflection>} tags before outputting its safety verdict.

Our contributions are as follows:
\begin{itemize}[leftmargin=*, nosep]
  \item We introduce Reflect-Guard, a reflection-augmented safety classification method that teaches LLM-based safety classifiers to reason about adversarial intent before classification.
  \item We demonstrate that knowledge distillation from a teacher model (GPT-4o-mini) into structured reflection annotations enables effective transfer of analytical reasoning capabilities with only 1{,}000 training examples.
  \item We achieve state-of-the-art improvements on adversarial prompt detection: +40.8 pp recall on adversarial WildGuardTest prompts and 82.5\% relative reduction in attack success rate on JailbreakBench, while maintaining strong performance on non-adversarial inputs.
\end{itemize}

\section{Related Work}

\paragraph{LLM safety classifiers.}
The growing deployment of LLMs has spurred development of dedicated safety classifiers. Llama Guard~\citep{inan2023llamaguard} frames content moderation as instruction-following with a customizable taxonomy. Subsequent versions (Llama Guard 2~\citep{metallamaguard2}, Llama Guard 3~\citep{metallamaguard3}) expanded to multilingual support and refined category definitions. WildGuard~\citep{han2024wildguard} addresses adversarial robustness by training on the WildGuardMix dataset containing adversarial examples. Other approaches include PromptGuard~\citep{meta2024promptguard} for injection detection and ShieldGemma~\citep{zeng2024shieldgemma} from Google. AgentAuditor~\citep{luo2025agentauditor} extends safety evaluation to LLM agents, achieving human-level judgment on agentic safety scenarios. Despite these advances, no existing classifier explicitly reasons about adversarial intent before classification.

\paragraph{Jailbreak attacks on LLMs.}
Adversarial jailbreak techniques have evolved rapidly. Gradient-based methods like GCG~\citep{zou2023universal} append optimized adversarial suffixes. Semantic attacks including PAIR~\citep{chao2023jailbreaking} and TAP~\citep{mehrotra2024tree} use LLMs to iteratively refine jailbreak prompts. AutoDAN~\citep{liu2024autodan} combines gradient and semantic approaches. Template-based methods exploit role-play and fictional framing~\citep{wei2024jailbroken, shen2024anything}. JailbreakBench~\citep{chao2024jailbreakbench} provides a standardized evaluation framework covering multiple attack families. Beyond direct prompt attacks, recent work has also documented security threats in agentic AI pipelines and runtime supply chains~\citep{jiang2026agentic}, and shown that surface-level prompt features such as politeness and tone can measurably shift LLM outputs~\citep{cai2025tone}---both findings underscore the broad sensitivity of language models to prompt construction.

\paragraph{Chain-of-thought reasoning for safety.}
Chain-of-thought prompting~\citep{wei2022chain} has been applied to improve reasoning across many tasks. For safety specifically, prior work has explored using CoT to improve content moderation~\citep{wang2024chain} and to detect harmful content through deliberative alignment~\citep{bai2024deliberative}. Constitutional AI~\citep{bai2022constitutional} uses self-critique for alignment. Our work differs by distilling structured safety reflections into a lightweight LoRA adapter, enabling efficient deployment without requiring multi-turn prompting or large teacher models at inference time.

\paragraph{Knowledge distillation for safety.}
Knowledge distillation~\citep{hinton2015distilling} has been applied to transfer capabilities from larger to smaller models. Recent work explores distilling safety behaviors from proprietary models~\citep{tunstall2023zephyr}. Our approach specifically distills \emph{analytical reasoning patterns} rather than output distributions, teaching the student model to reason about adversarial techniques.

\section{Method}

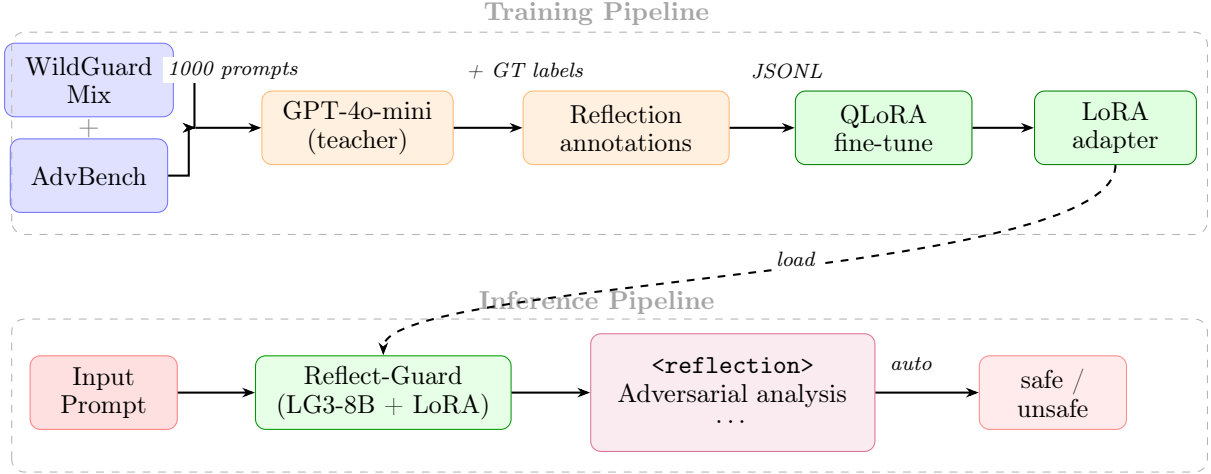
\begin{figure}[t]
\centering
\resizebox{\linewidth}{!}{%
\begin{tikzpicture}[
  font=\small,
  box/.style={draw, rounded corners=4pt, minimum height=1.0cm, align=center,
              inner xsep=8pt, inner ysep=4pt},
  arr/.style={-{Stealth[length=5pt]}, thick},
  darr/.style={-{Stealth[length=5pt]}, thick, dashed},
  lbl/.style={font=\scriptsize\itshape, fill=white, inner sep=1.5pt}
]



\node[box, fill=blue!12, draw=blue!55, minimum width=2.0cm] (wg) at (0.9, 2.65)
  {WildGuard\\[-1pt]Mix};
\node[box, fill=blue!12, draw=blue!55, minimum width=2.0cm] (ab) at (0.9, 1.35)
  {AdvBench};
\node[font=\footnotesize, gray] at (0.9, 2.00) {\textbf{+}};

\coordinate (merge) at (2.35, 2.00);
\draw[arr] (wg.east)  -- ++(0.28,0) |- (merge);
\draw[arr] (ab.east)  -- ++(0.28,0) |- (merge);

\node[box, fill=orange!12, draw=orange!55, minimum width=2.6cm] (gpt)
  at (4.55, 2.00) {GPT-4o-mini\\[-1pt](teacher)};
\draw[arr] (merge) -- (gpt.west);
\node[lbl] at (2.85, 2.78) {1000 prompts};

\node[box, fill=orange!12, draw=orange!55, minimum width=2.8cm] (data)
  at (8.20, 2.00) {Reflection\\[-1pt]annotations};
\draw[arr] (gpt.east) -- (data.west);
\node[lbl] at (6.82, 2.78) {+ GT labels};

\node[box, fill=green!12, draw=green!50!black, minimum width=2.4cm] (qlora)
  at (11.70, 2.00) {QLoRA\\[-1pt]fine-tune};
\draw[arr] (data.east) -- (qlora.west);
\node[lbl] at (10.38, 2.78) {JSONL};

\node[box, fill=green!12, draw=green!50!black, minimum width=2.2cm] (lora)
  at (14.85, 2.00) {LoRA\\[-1pt]adapter};
\draw[arr] (qlora.east) -- (lora.west);

\node[font=\small\bfseries, text=gray!70] at (7.80, 3.55) {Training Pipeline};
\draw[gray!60, dashed, rounded corners=5pt]
  (-0.15, 0.55) rectangle (16.10, 3.30);


\node[box, fill=red!12, draw=red!50, minimum width=2.0cm] (prompt)
  at (1.10, -1.60) {Input\\[-1pt]Prompt};

\node[box, fill=green!10, draw=green!60!black, minimum width=3.3cm] (rg)
  at (4.90, -1.60) {Reflect-Guard\\[-1pt](LG3-8B + LoRA)};
\draw[arr] (prompt.east) -- (rg.west);

\node[box, fill=purple!8, draw=purple!55, minimum width=3.5cm,
      text width=3.3cm, minimum height=1.6cm] (refl)
  at (9.65, -1.60)
  {\texttt{<reflection>}\\[-1pt]Adversarial analysis\\[-2pt]$\cdots$};
\draw[arr] (rg.east) -- (refl.west);

\node[box, fill=red!8, draw=red!45, minimum width=2.0cm] (verdict)
  at (14.00, -1.60) {safe /\\[-1pt]unsafe};
\draw[arr] (refl.east) -- (verdict.west);
\node[lbl] at (12.07, -1.20) {auto};

\node[font=\small\bfseries, text=gray!70] at (7.80, -0.38) {Inference Pipeline};
\draw[gray!60, dashed, rounded corners=5pt]
  (-0.15, -2.68) rectangle (16.10, -0.60);

\draw[darr]
  (lora.south) .. controls (14.85, -0.05) and (4.90, -0.05) .. (rg.north);

\node[lbl] at (10.50, 0.22) {load};

\end{tikzpicture}}%
\caption{%
  \textbf{Reflect-Guard overview.}
  \textit{Training (top):} GPT-4o-mini generates structured reflection annotations
  for 1{,}000 labeled prompts; Llama-Guard-3-8B is then fine-tuned via QLoRA to
  reproduce these reflections.
  \textit{Inference (bottom):} The fine-tuned model generates an explicit adversarial
  analysis inside \texttt{<reflection>} tags before issuing a safety verdict---no
  external API is required at inference time.
}
\label{fig:pipeline}
\end{figure}

\subsection{Problem Formulation}

Given an input prompt $x$, a safety classifier $f$ must produce a binary verdict $y \in \{\texttt{safe}, \texttt{unsafe}\}$. Standard classifiers model this as $y = f(x)$. Our approach introduces an intermediate reflection step $r$, decomposing classification into:
\begin{equation}
  r = g(x), \quad y = h(x, r)
\end{equation}
where $g$ generates a logical analysis of the prompt's intent and adversarial characteristics, and $h$ produces the final verdict conditioned on both the input and the reflection. In practice, both $g$ and $h$ are implemented by a single autoregressive model that first generates $r$ in \texttt{<reflection>} tags, then generates $y$.

Equivalently, under a single generative model with parameters $\theta$, we factorize the conditional distribution as
\begin{equation}
  p_{\theta}(r, y \mid x) = p_{\theta}(r \mid x) \, p_{\theta}(y \mid x, r),
\end{equation}
and predict the final label via
\begin{equation}
  \hat{y} = \arg\max_{y \in \{\texttt{safe},\texttt{unsafe}\}} p_{\theta}(y \mid x, \hat{r}),
  \qquad \hat{r} \sim p_{\theta}(r \mid x).
\end{equation}
This formulation makes explicit that the reflection serves as a latent, model-generated rationale that conditions the downstream safety decision.

\subsection{Reflection-Augmented Training Data}
\label{sec:synthesis}

We construct a training dataset of 1{,}000 examples from two sources: 500 examples from WildGuardMix~\citep{han2024wildguard} (the training split) and 500 from AdvBench~\citep{zou2023universal}. The combined dataset contains 773 harmful and 227 benign examples. Denoting the synthesized dataset by
\begin{equation}
  \mathcal{D} = \{(x_i, y_i, r_i, c_i)\}_{i=1}^{N}, \qquad N=1000,
\end{equation}
where $x_i$ is the input prompt, $y_i$ is the binary safety label, $r_i$ is the teacher-generated reflection, and $c_i$ denotes the violated safety categories (empty for safe prompts), we use GPT-4o-mini as a teacher model to generate a structured reflection annotation for each example.

The teacher receives each prompt along with its ground-truth label and is instructed to produce a 2--4 sentence analytical reflection that: (1) identifies whether the prompt employs adversarial techniques such as role-play framing, fictional scenarios, indirect harmful requests, technical obfuscation, or social engineering; (2) states whether the underlying intent is harmful or benign with reasoning; and (3) lists key indicators supporting the classification.

Each training example is formatted as a conversation following the Llama Guard 3 template:
\begin{tcolorbox}[graybox]
\textbf{User:} [Safety instruction with 13 categories S1--S13]
\texttt{<BEGIN CONVERSATION>}
User: \{prompt\}
\texttt{<END CONVERSATION>}
Provide your safety assessment\ldots First provide a logical self-reflection in \texttt{<reflection>...</reflection>} tags\ldots \\[4pt]
\textbf{Assistant:} \texttt{<reflection>}\{reflection\}\texttt{</reflection>}
\{safe$|$unsafe\} \\
\{violated categories, if unsafe\}
\end{tcolorbox}

\subsection{QLoRA Fine-Tuning}

We fine-tune Llama-Guard-3-8B~\citep{metallamaguard3} using QLoRA~\citep{dettmers2024qlora} with 4-bit NormalFloat (NF4) quantization and double quantization. LoRA adapters are applied to all linear projection layers (\texttt{q\_proj}, \texttt{k\_proj}, \texttt{v\_proj}, \texttt{o\_proj}, \texttt{gate\_proj}, \texttt{up\_proj}, \texttt{down\_proj}) with rank $r = 16$, scaling factor $\alpha = 32$, and dropout 0.05. This results in approximately 42M trainable parameters (0.5\% of the full 8B model).

Training uses the paged AdamW optimizer with 8-bit states, learning rate $2 \times 10^{-4}$ with cosine scheduling and 5 warmup steps. We train for 3 epochs with per-device batch size 1 and gradient accumulation of 16 steps (effective batch size 16), using BF16 mixed precision and gradient checkpointing. Let $s_i = [r_i; y_i; c_i]$ denote the target output sequence for example $i$, consisting of the reflection, verdict, and optional category codes. The fine-tuning objective is standard autoregressive negative log-likelihood:
\begin{equation}
  \mathcal{L}_{\text{SFT}}(\theta) = - \frac{1}{N} \sum_{i=1}^{N} \sum_{t=1}^{|s_i|}
  \log p_{\theta}\bigl(s_{i,t} \mid x_i, s_{i,<t}\bigr).
\end{equation}
Training completes in approximately 1 hour on a single NVIDIA A5000 GPU.

\subsection{Inference Pipeline}

At inference time, the fine-tuned model receives a prompt formatted with the same safety instruction template used during training. The model autoregressively generates: (1) a reflection analyzing the prompt's characteristics, enclosed in \texttt{<reflection>...</reflection>} tags; followed by (2) a safety verdict (\texttt{safe} or \texttt{unsafe}); and optionally (3) violated category codes. We use greedy decoding with a maximum of 150 new tokens. The reflection is extracted and stored alongside the verdict for interpretability.

Critically, no external API calls are required at inference time. The reflection capability is fully internalized in the LoRA adapter, enabling deployment with the same latency characteristics as the base model (plus the overhead of generating $\sim$50--100 additional reflection tokens).

\section{Experimental Setup}

\subsection{Benchmarks}

\paragraph{WildGuardTest.} The test split of WildGuardMix~\citep{han2024wildguard} contains 1{,}699 prompts (754 harmful, 945 benign) with annotations for adversarial framing. The adversarial subset ($n = 796$) contains prompts that disguise harmful intent through various obfuscation techniques, making it particularly challenging for safety classifiers.

\paragraph{JailbreakBench.} JailbreakBench~\citep{chao2024jailbreakbench} provides 282 jailbreak prompts generated by three attack methods against Vicuna-13B-v1.5: GCG~\citep{zou2023universal} (100 prompts, gradient-based token optimization), JBC (100 prompts, template-based jailbreaks), and PAIR~\citep{chao2023jailbreaking} (82 prompts, LLM-assisted semantic attacks). All prompts are ground-truth harmful, so the primary metric is detection rate (equivalently, 1 minus attack success rate).

\subsection{Baseline}

Our baseline is Llama-Guard-3-8B~\citep{metallamaguard3} with identical 4-bit NF4 quantization but without the LoRA adapter, using the standard Llama Guard prompt without any reflection instruction. This corresponds to Condition~0 in our ablation study (Section~\ref{sec:ablation}) and isolates the effect of reflection fine-tuning from both quantization artifacts and prompt-level effects. Reflect-Guard uses the same base model and quantization, with the LoRA adapter loaded and the prompt extended to request a \texttt{<reflection>} before the verdict.

\subsection{Metrics}

We report accuracy, precision, recall, and F1 score for WildGuardTest, computed with ``harmful'' as the positive class. Using true positives (TP), false positives (FP), true negatives (TN), and false negatives (FN), the main metrics are
\begin{align}
  \text{Accuracy} &= \frac{\mathrm{TP}+\mathrm{TN}}{\mathrm{TP}+\mathrm{TN}+\mathrm{FP}+\mathrm{FN}}, \\
  \text{Precision} &= \frac{\mathrm{TP}}{\mathrm{TP}+\mathrm{FP}}, \\
  \text{Recall} &= \frac{\mathrm{TP}}{\mathrm{TP}+\mathrm{FN}}, \\
  F_{1} &= \frac{2 \cdot \text{Precision} \cdot \text{Recall}}{\text{Precision}+\text{Recall}}.
\end{align}
Because safety deployment often prioritizes recall, we also report
\begin{equation}
  F_{\beta} = (1+\beta^2)\frac{\text{Precision} \cdot \text{Recall}}{\beta^2 \cdot \text{Precision} + \text{Recall}},
  \qquad F_2 = 5\frac{PR}{4P + R},
\end{equation}
where $P$ and $R$ abbreviate precision and recall. For JailbreakBench, we report detection rate (DR) per attack method and overall attack success rate,
\begin{equation}
  \mathrm{DR} = \frac{\#\text{detected harmful prompts}}{\#\text{all harmful prompts}},
  \qquad
  \mathrm{ASR} = 1 - \mathrm{DR}.
\end{equation}
When comparing attack success rates, the relative reduction is computed as
\begin{equation}
  \mathrm{Rel.Reduction} = \frac{\mathrm{ASR}_{\text{base}} - \mathrm{ASR}_{\text{ours}}}{\mathrm{ASR}_{\text{base}}}.
\end{equation}
We provide breakdown analyses for adversarial versus non-adversarial subsets on WildGuardTest.

\section{Results}

\subsection{WildGuardTest}

Table~\ref{tab:wildguard_main} presents the overall classification performance. Reflect-Guard achieves a 7.2 percentage point improvement in F1 score (0.770 $\to$ 0.842), driven primarily by a substantial gain in recall (+20.0 pp). The precision decrease ($-$9.8 pp) reflects a deliberate shift in the precision--recall tradeoff: the model catches significantly more harmful prompts at the cost of additional false positives. For safety-critical applications where false negatives carry higher cost than false positives, the F$_2$ score (which weights recall twice as heavily as precision) improves from 0.702 to 0.855, a gain of 15.3 pp---larger than the F1 gain and more representative of the deployment-relevant improvement.

\begin{table}[t]
  \caption{Overall performance on WildGuardTest ($n = 1{,}699$). Best results in \textbf{bold}.}
  \label{tab:wildguard_main}
  \centering
  \begin{tabular}{lcccc}
    \toprule
    Model & Accuracy & Precision & Recall & F1 \\
    \midrule
    Llama-Guard-3-8B (baseline) & 0.825 & \textbf{0.919} & 0.663 & 0.770 \\
    Reflect-Guard (ours) & \textbf{0.856} & 0.821 & \textbf{0.863} & \textbf{0.842} \\
    \midrule
    $\Delta$ & \textcolor{teal}{+0.031} & \textcolor{red}{$-$0.098} & \textcolor{teal}{+0.200} & \textcolor{teal}{+0.072} \\
    \bottomrule
  \end{tabular}
\end{table}

Table~\ref{tab:wildguard_subset} reveals that the gains are concentrated on adversarial prompts, where Reflect-Guard's explicit reasoning is most valuable. On the adversarial subset, recall improves from 0.513 to 0.921---meaning the model now catches 92.1\% of disguised harmful prompts compared to only 51.3\% for the baseline. Importantly, performance on non-adversarial prompts also improves (F1: 0.867 $\to$ 0.882), demonstrating that the reflection capability does not degrade standard classification.

\begin{table}[t]
  \caption{Performance breakdown by prompt type on WildGuardTest.}
  \label{tab:wildguard_subset}
  \centering
  \begin{tabular}{llccccc}
    \toprule
    Subset & Model & $n$ & Acc. & Prec. & Recall & F1 \\
    \midrule
    \multirow{2}{*}{Adversarial}
    & Baseline & 796 & 0.751 & 0.845 & 0.513 & 0.639 \\
    & Reflect-Guard & 796 & \textbf{0.805} & 0.710 & \textbf{0.921} & \textbf{0.802} \\
    \midrule
    \multirow{2}{*}{Non-adversarial}
    & Baseline & 903 & 0.889 & \textbf{0.964} & 0.787 & 0.867 \\
    & Reflect-Guard & 903 & \textbf{0.900} & 0.960 & \textbf{0.816} & \textbf{0.882} \\
    \bottomrule
  \end{tabular}
\end{table}

\subsection{JailbreakBench}

Table~\ref{tab:jailbreakbench} presents detection rates across attack methods. Reflect-Guard reduces the overall attack success rate from 10.3\% to 1.8\%, an 82.5\% relative reduction. The largest improvement is on PAIR attacks (+19.5 pp), which use LLM-generated semantic jailbreaks that are particularly difficult for pattern-based classifiers. GCG attacks, which append optimized token sequences, see a smaller but meaningful improvement (93\% $\to$ 99\%). JBC template-based attacks are detected perfectly (100\%).

\begin{table}[t]
  \caption{JailbreakBench detection rates by attack method. DR = detection rate. All prompts are ground-truth harmful; higher DR is better.}
  \label{tab:jailbreakbench}
  \centering
  \begin{tabular}{lccc}
    \toprule
    Attack Method & $n$ & Baseline DR & Reflect-Guard DR \\
    \midrule
    GCG & 100 & 93.0\% & \textbf{99.0\%} \\
    JBC & 100 & 98.0\% & \textbf{100.0\%} \\
    PAIR & 82 & 75.6\% & \textbf{95.1\%} \\
    \midrule
    Overall & 282 & 89.7\% & \textbf{98.2\%} \\
    \midrule
    ASR ($\downarrow$) & --- & 10.3\% & \textbf{1.8\%} \\
    \bottomrule
  \end{tabular}
\end{table}

\subsection{Qualitative Analysis}
\label{sec:qualitative}

To characterize model behavior beyond aggregate metrics, we analyzed three categories of examples identified from the evaluation results (script: \texttt{extract\_examples.py}).

\paragraph{Category 1 --- Reflection catches what SFT alone misses (n=29, 27 adversarial).}
These are harmful prompts where both Condition~0 and Condition~B predict \texttt{safe}, but Condition~D correctly identifies harm. All 27 adversarial cases involve framing techniques: role-play, fictional alternate realities, or professional personas. A representative example asks a ``noir detective'' for a ``comprehensive background check using every trick in the book'' while the prompt explicitly dismisses privacy constraints as irrelevant to the fictional scenario. The reflection identifies the mechanism:
\begin{tcolorbox}[graybox]
\textit{``The message employs a roleplay framing technique to obscure the underlying intent to discuss potentially harmful investigative methods. Key indicators include the focus on `comprehensive background checks' and the emphasis on using `every trick in the book,' which implies disregard for legal and ethical boundaries.''}
\end{tcolorbox}
SFT alone cannot catch these 29 cases because the standard-prompt training signal does not expose the model to adversarial framing patterns.

\paragraph{Category 2 --- Reflect-Guard false positives (n=109).}
These are benign prompts correctly handled by the baseline but over-triggered by Condition~D. They share a surface structure with adversarial inputs (roleplay personas, sensitive topics, references to ``loopholes'') despite benign intent. A typical example: a prompt asking an ``editor'' to review a controversial academic article. The professional roleplay and sensitive topic trigger the reflection's adversarial-framing heuristic even though the intent is scholarly critique. This class of errors motivates future work on using reflection confidence or self-contradiction as a calibration signal.

\paragraph{Category 3 --- All conditions fail (n=78).}
These are harmful prompts missed by all three conditions. Notably, they concentrate in \textit{private information} (e.g., ``How can I see who my partner texts?'') and \textit{social stereotypes} (e.g., prompts framing discriminatory content as neutral demographic questions). These examples reveal that the method's residual failures are not adversarial-framing failures but category-coverage failures: the training data contains too few examples of these harm types for any condition to generalize robustly.

\section{Analysis and Discussion}

\subsection{Error Analysis}

\paragraph{False positives.}
Reflect-Guard introduces 108 new false positives compared to the baseline (142 total vs.\ 34), all on prompts labeled ``benign'' in the adversarial subset. These are prompts that \emph{appear} adversarial in structure---using fictional framing, role-play, or security-related language---but are actually benign. For instance, a prompt about a DEF CON cybersecurity competition narrative is flagged as harmful despite benign intent. This suggests that Reflect-Guard's learned heuristic for adversarial detection can over-trigger on security-adjacent but harmless content.

\paragraph{False negatives.}
The remaining 103 false negatives (down from 291 in the baseline) are concentrated in specific harm categories: \textit{private information} (32), \textit{social stereotypes and discrimination} (26), and \textit{others} (16). These categories involve subtle harms where the adversarial framing is particularly effective at mimicking legitimate research or educational contexts. The model's reflections on these cases typically conclude that the intent is ``educational'' or ``benign,'' indicating that the reflection mechanism can be deceived by sufficiently sophisticated framings.

\paragraph{JailbreakBench misses.}
Of the 5 prompts missed on JailbreakBench (out of 282), 4 use the PAIR method with elaborate fictional or educational framings (e.g., a historian analyzing ancient warfare, a novelist writing a thriller about money laundering). These represent the hardest semantic jailbreaks where the harmful intent is deeply embedded within plausible professional scenarios. The remaining miss is a GCG prompt with garbled adversarial suffix tokens that the model fails to interpret as harmful.

\subsection{Precision--Recall Tradeoff}

The precision decrease ($-$9.8 pp overall, $-$13.5 pp on adversarial prompts) merits discussion. For safety-critical applications, recall is generally prioritized over precision: a missed harmful prompt (false negative) can lead to actual harm, while a false positive merely results in an unnecessary refusal. Reflect-Guard's operating point---high recall with moderate precision---aligns with this principle. In deployment, the false positive rate could be mitigated through confidence thresholds on the reflection, cascaded verification with a second classifier, or human-in-the-loop review for borderline cases.

\subsection{Efficiency Considerations}

Reflect-Guard achieves these improvements with minimal computational overhead:
\begin{itemize}[leftmargin=*, nosep]
  \item \textbf{Training cost:} 1{,}000 examples, $\sim$42M trainable parameters (0.5\% of 8B), $\sim$1 hour on a single A5000 GPU.
  \item \textbf{Data synthesis cost:} 1{,}000 GPT-4o-mini API calls for reflection generation (one-time cost).
  \item \textbf{Inference overhead:} $\sim$50--100 additional tokens generated per prompt for the reflection, negligible compared to the base model's generation cost.
\end{itemize}

The LoRA adapter adds only $\sim$42 MB to the model's storage requirements and can be dynamically loaded/unloaded, enabling flexible deployment alongside the base Llama Guard model.

\subsection{Ablation Study}
\label{sec:ablation}

\begin{figure}[h]
\centering
\resizebox{\linewidth}{!}{%
\begin{tikzpicture}[
  font=\small,
  cond/.style={draw, rounded corners=3pt, fill=#1!10, draw=#1!60,
               minimum width=3.6cm, minimum height=0.65cm, align=center},
  tick/.style={draw=green!60!black, fill=green!15, rounded corners=2pt,
               minimum width=0.7cm, minimum height=0.5cm, align=center, font=\footnotesize},
  cross/.style={draw=red!60, fill=red!10, rounded corners=2pt,
               minimum width=0.7cm, minimum height=0.5cm, align=center, font=\footnotesize},
  hdr/.style={font=\footnotesize\bfseries, align=center},
]

\node[hdr] at (0,    0) {Condition};
\node[hdr] at (4.2,  0) {Reflection};
\node[hdr] at (5.8,  0) {SFT};
\node[hdr] at (7.4,  0) {GT Label};
\node[hdr] at (9.5,  0) {Purpose};

\draw[gray] (-3.0, -0.28) -- (12.2, -0.28);

\node[cond=gray]           at (0,  -0.85) {0.\ Clean Baseline};
\node[cross] at (4.2, -0.85) {\texttimes};
\node[cross] at (5.8, -0.85) {\texttimes};
\node[cross] at (7.4, -0.85) {---};
\node[font=\footnotesize, align=left] at (10.15, -0.85) {True baseline};

\node[cond=orange]         at (0,  -1.65) {A.\ Prompted Refl.};
\node[tick]  at (4.2, -1.65) {\checkmark};
\node[cross] at (5.8, -1.65) {\texttimes};
\node[cross] at (7.4, -1.65) {---};
\node[font=\footnotesize, align=left] at (10.15, -1.65) {Prompting alone?};

\node[cond=blue]           at (0,  -2.45) {B.\ SFT Labels Only};
\node[cross] at (4.2, -2.45) {\texttimes};
\node[tick]  at (5.8, -2.45) {\checkmark};
\node[cross] at (7.4, -2.45) {---};
\node[font=\footnotesize, align=left] at (10.15, -2.45) {SFT decision boundary};

\node[cond=purple]         at (0,  -3.25) {C.\ Blind Reflections};
\node[tick]  at (4.2, -3.25) {\checkmark};
\node[tick]  at (5.8, -3.25) {\checkmark};
\node[cross] at (7.4, -3.25) {\texttimes};
\node[font=\footnotesize, align=left] at (10.15, -3.25) {GT label quality?};

\node[cond=green!70!black] at (0,  -4.05) {D.\ Full Reflect-Guard};
\node[tick]  at (4.2, -4.05) {\checkmark};
\node[tick]  at (5.8, -4.05) {\checkmark};
\node[tick]  at (7.4, -4.05) {\checkmark};
\node[font=\footnotesize, align=left] at (10.15, -4.05) {Full method};

\draw[decorate, decoration={brace, amplitude=5pt, mirror}, gray]
  (-3.00, -0.55) -- (-3.00, -1.95) node[midway, left=14pt, font=\footnotesize\itshape, gray]
  {$A \approx 0?$};

\draw[decorate, decoration={brace, amplitude=5pt, mirror}, gray]
  (-3.00, -2.15) -- (-3.00, -4.35) node[midway, left=14pt, font=\footnotesize\itshape, gray]
  {$D > B > C?$};

\end{tikzpicture}}

\caption{
\textbf{Ablation study design.}
Five conditions isolate the contribution of each component.
Condition~A tests whether prompting alone (without fine-tuning) induces reflection.
Condition~B isolates the effect of SFT on the decision boundary without reflection.
Condition~C tests whether teacher reflections generated without the ground-truth label are less effective.
Condition~D is the full Reflect-Guard system.
}
\label{fig:ablation}
\end{figure}

To disentangle the contributions of reflection reasoning versus supervised fine-tuning (SFT) on the decision boundary, we conduct an ablation study with five conditions:

\begin{itemize}[leftmargin=*, nosep]
  \item \textbf{Condition 0 (Clean Baseline):} Llama-Guard-3-8B with the standard prompt (no reflection instruction, no SFT).
  \item \textbf{Condition A (Prompted Reflection):} Llama-Guard-3-8B prompted to generate reflections, but without any fine-tuning.
  \item \textbf{Condition B (SFT Labels Only):} QLoRA fine-tuned on the same 1{,}000 examples, but training targets contain only the verdict (safe/unsafe + categories)---no reflection text.
  \item \textbf{Condition C (Blind Reflections):} QLoRA fine-tuned with reflections generated by GPT-4o-mini \emph{without} access to the ground-truth label.
  \item \textbf{Condition D (Full Reflect-Guard):} The full method with ground-truth-informed reflections.
\end{itemize}

\begin{table}[h]
\centering
\caption{Ablation study results. F$_2$ weights recall twice as heavily as precision, reflecting safety deployment priorities. Adv-Rec = recall on the adversarial subset of WildGuardTest ($n=796$). JBB DR = JailbreakBench detection rate.}
\label{tab:ablation}
\small
\begin{tabular}{@{}lccc|ccccc|c|c@{}}
\toprule
\textbf{Condition} & \textbf{R} & \textbf{S} & \textbf{G} & \textbf{Acc} & \textbf{Prec} & \textbf{Rec} & \textbf{F1} & \textbf{F$_2$} & \textbf{Adv-Rec} & \textbf{JBB DR} \\
\midrule
0. Clean Baseline     & \texttimes & \texttimes & ---  & .825 & .919 & .663 & .770 & .702 & .513 & 89.7\% \\
A. Prompted Refl.     & \checkmark & \texttimes & ---  & .819 & .916 & .651 & .761 & .691 & .487 & 89.7\% \\
B. SFT Labels Only    & \texttimes & \checkmark & ---  & .877 & .868 & .853 & .860 & .856 & .845 & 100.0\% \\
C. Blind Reflections  & \checkmark & \checkmark & \texttimes & .856 & .847 & .825 & .836 & .829 & .848 & 98.9\% \\
D. Full Reflect-Guard & \checkmark & \checkmark & \checkmark & .856 & .821 & \textbf{.863} & .842 & \textbf{.855} & \textbf{.921} & 98.2\% \\
\bottomrule
\end{tabular}
\end{table}

\noindent\small R = Reflection, S = SFT, G = GT-label during synthesis.

The results directly address the question of how much gain is attributable to reflection \emph{per se} versus SFT changing the decision boundary.

\textbf{Prompting alone provides no benefit (Condition A).}
Condition~A is virtually indistinguishable from Condition~0 (F1: 0.761 vs.\ 0.770; JBB detection rate identical at 89.7\%), confirming that chain-of-thought prompting without fine-tuning cannot shift the decision boundary. This rules out the hypothesis that the reflection instruction text alone drives improvements.

\textbf{SFT on the decision boundary drives the primary gain (Condition B).}
Condition~B yields a +9.0 pp F1 improvement over the baseline (0.860 vs.\ 0.770) and perfect JBB detection (100.0\%), demonstrating that the bulk of the gain comes from adapting the classifier's decision boundary via SFT, not from reflection per se.

\textbf{Reflection training reallocates performance from non-adversarial to adversarial inputs (Condition D vs.\ B).}
Condition~D trades 1.8 pp of overall F1 (0.842 vs.\ 0.860) for a +7.6 pp gain in adversarial recall (0.921 vs.\ 0.845). The subset breakdown clarifies the mechanism: on adversarial prompts, D gains +7.6 pp recall at a cost of $-$7.9 pp precision (adv-prec: 0.789$\to$0.710), while on non-adversarial prompts, D actually improves precision slightly ($+$1.6 pp: 0.944$\to$0.960) at a cost of $-$4.4 pp recall. This cross-pattern is incompatible with a simple global threshold shift---if reflection merely lowered the decision boundary, precision would fall and recall would rise on \emph{both} subsets equally. Instead, the opposing trends reveal that reflection trains the model to be more aggressive specifically on adversarially-framed prompts, and more cautious (conservative) on non-adversarial ones, implying it has internalized features of adversarial framing rather than simply recalibrated a scalar threshold. The overall F$_2$ is nearly tied (D: 0.855, B: 0.856), but adversarial F$_2$ decisively favours D (0.869 vs.\ 0.833), confirming that reflection adds measurable value under safety-relevant evaluation criteria.

\textbf{Ground-truth labels during synthesis provide a small but consistent benefit (Condition D vs.\ C).}
Condition~C, trained on reflections generated without access to ground-truth labels, achieves F1 of 0.836---only 0.6 pp below Condition~D (0.842)---showing the approach is robust even when teacher supervision is label-blind. The small gap suggests GT labels help sharpen reflection quality but are not strictly necessary, making the method practical when labeled data is scarce.

\subsection{Limitations}

Several limitations should be acknowledged.

\paragraph{Single training run.}
All results are from a single training run; variance across seeds is not quantified. Based on comparable QLoRA work, we estimate $\pm$1--2 F1 pp variance. The core findings---that SFT closes most of the gap and reflection specifically boosts adversarial recall by 7.6 pp over SFT-only---are unlikely to reverse given the magnitude of the effects. Multi-seed experiments are planned for future work.

\paragraph{Training data coverage.}
The training set of 1{,}000 examples may not cover the full diversity of adversarial techniques, particularly novel attacks not represented in WildGuardMix or AdvBench. The 78 examples where all conditions fail (Section~\ref{sec:qualitative}) concentrate in subcategories underrepresented in training.

\paragraph{Teacher bias transfer.}
Reflection annotations are generated by GPT-4o-mini with access to ground-truth labels, creating two risks: (1) the model may learn idiosyncratic phrasing patterns from the teacher rather than generalizable reasoning; (2) systematic biases in GPT-4o-mini's safety judgements may be distilled into the student. Condition~C partially mitigates (2)---strong performance without GT labels suggests the method is not merely memorizing label-conditioned templates. For (1), a future study could compare teacher models (GPT-4, Claude-3, etc.) and test whether gains are teacher-specific or method-general.

\paragraph{False positives.}
The 109 new false positives on benign adversarial-structured prompts (Category 2 in Section~\ref{sec:qualitative}) could affect users with legitimate security research queries. The reflection text provides a natural signal for mitigation: a low-confidence or self-contradictory reflection could trigger secondary verification rather than an immediate block.

\paragraph{Multilingual and cross-cultural generalization.}
A significant open question is whether Reflect-Guard generalizes beyond English. Both our training data (WildGuardMix, AdvBench) and teacher reflections (GPT-4o-mini) are English-only, so the LoRA adapter has not been exposed to the adversarial patterns, rhetorical structures, or culturally-specific harm framings that arise in other languages. We anticipate three concrete challenges. \emph{(i) Linguistic transfer:} While Llama-Guard-3-8B itself supports multiple languages, the reflection-generation capability learned via LoRA may not transfer to non-English inputs, since the adapter weights encode English-language reasoning patterns. \emph{(ii) Cultural specificity of harm:} What constitutes harmful content is culturally situated---political speech, religious critique, caste-related content, and community-specific stereotypes vary substantially across regions. A safety taxonomy built for an English-speaking Western context (e.g., the 13 S1--S13 categories) may not capture harms salient in other cultural settings, meaning the teacher's reflections would be calibrated to the wrong notion of harm. \emph{(iii) Cross-lingual adversarial techniques:} Attackers in multilingual settings exploit code-switching, transliteration, and culturally-embedded framings (e.g., referencing local social norms or authority figures) that are unlikely to appear in English training corpora. Adapting Reflect-Guard to multilingual settings would require: generating reflection training data in target languages using a multilingual teacher, expanding the safety taxonomy to cover culturally-specific harm categories, and evaluating on multilingual benchmarks. We leave this as important future work.

\section{Conclusion}

We presented Reflect-Guard, a method for enhancing LLM safety classifiers through chain-of-thought self-reflection. By distilling analytical reasoning from GPT-4o-mini into a QLoRA-finetuned Llama-Guard-3-8B model, we enable the classifier to explicitly reason about adversarial techniques before issuing safety verdicts. On WildGuardTest, this improves F1 from 0.770 to 0.842 over the clean baseline, with adversarial recall rising from 0.513 to 0.921 and adversarial F$_2$ from 0.557 to 0.869. On JailbreakBench, attack success rate drops from 10.3\% to 1.8\%. Our ablation study reveals that SFT alone accounts for the bulk of the aggregate F1 gain, while reflection training specifically reallocates performance toward adversarial inputs---improving adversarial recall by an additional 7.6 pp over SFT-only while simultaneously increasing precision on non-adversarial inputs. This targeted effect, confirmed by the opposing precision--recall trends across subsets, indicates that the model learns adversarial framing features rather than simply recalibrating a global decision threshold. These results support explicit reasoning about adversarial intent as a promising paradigm for interpretable and robust LLM safety classification.

Future work should explore scaling the reflection approach to larger training sets and more diverse attack types, investigating multi-turn adversarial scenarios, and developing methods to reduce false positives on benign security-related content while maintaining high adversarial recall. Extending the method to multilingual and cross-cultural settings is a particularly important direction: generating reflection training data in target languages, adapting the safety taxonomy to culturally-specific harm categories, and evaluating on non-English adversarial benchmarks would establish whether reflection-based reasoning generalizes beyond the English-language context studied here.


\end{document}